\documentclass[10pt, conference, letterpaper]{IEEEtran}
\usepackage{amsmath,amssymb,amsfonts,amsthm}
\usepackage{graphicx}
\usepackage{epstopdf}

\DeclareMathOperator*{\mini}{minimize}

\begin{document}
\title{Compressed Sensing-based Pilot Assignment and Reuse for Mobile UEs in mmWave Cellular Systems}
\author{
\authorblockN{Weng Chon Ao$^{\dag}$, Chenwei Wang$^*$, Ozgun Y. Bursalioglu$^*$, Haralabos Papadopoulos$^*$}
\authorblockA{$^{\dag}$University of Southern California, Los Angeles, CA 90089}
{\normalsize E-mail~:~wao@usc.edu}
\authorblockA{$^*$DOCOMO Innovations Inc., Palo Alto, CA 94304}
{\normalsize
E-mail~:~\{cwang, obursalioglu, hpapadopoulos\}@docomoinnovations.com}}

\maketitle

\begin{abstract}
Technologies for mmWave communication  are at the forefront of investigations in both industry and academia, as the mmWave band  offers the promise of orders of magnitude additional available bandwidths to what has already been allocated to cellular networks. The much larger number of antennas that can be supported in a small footprint at mmWave bands  can be leveraged to harvest massive-MIMO type beamforming and spatial multiplexing gains. Similar to LTE systems, two prerequisites for harvesting these benefits are detecting users and acquiring user channel state information (CSI) in the training phase. However, due to the fact that mmWave channels encounter much harsher propagation and decorrelate much faster, the tasks of user detection and CSI acquisition are both imperative and much more challenging than in LTE bands.

In this paper, we investigate the problem of fast user detection and CSI acquisition in the downlink of small cell mmWave networks. We assume TDD operation and  channel-reciprocity based CSI acquisition. To achieve densification benefits we propose pilot designs and channel estimators that leverage a combination of aggressive pilot reuse with fast user detection at the base station and compressed sensing channel estimation.  As our simulations show, the number of users that can be simultaneously served by the entire mmWave-band network with the proposed schemes increases substantially with respect to traditional compressed sensing based approaches with conventional pilot reuse.

%

\end{abstract}

\section{Introduction}

Owing to the prevalence of smart devices, the rapid growth of social networks, online video demands and internet of things, the data traffic conveyed by mobile communication networks has been soaring. As cellular communication systems continue to evolve, the design of the current mobile networks has been highly constrained by the scarcity of the radio spectrum. As a result, in the forthcoming 5G era, networks will be required to deliver large improvements in throughputs per unit area, user peak rates, massive device connectivity, significantly lower end-to-end latencies and lower energy costs. It is expected that the benefits of 5G will stem from both new radio access technologies and a new network infrastructure. It is widely accepted that to achieve the 5G dream of large throughput gains per unit area requires a combination of  additional bandwidth, network densification and technologies that offer spectral efficiency gains.

Massive MIMO, also known as ``Large-Scale" or ``Full-Dimension MIMO" was originally introduced by Marzatta \cite{Marzetta:2006, Marzetta:2010}. It  can provide large spectral efficiency gains through the use of a large number of antennas at the base stations (BSs). Compared to conventional MIMO, massive MIMO is usually associated with settings where the number of antennas at the BS is at least an order of magnitude larger than the number of users that are simultaneously served by the BS. The BS with massive MIMO is able to create very sharp beams to its users nearby, so as to shed more signal power on the desired users and less interference on undesired users. Also, due to channel hardening, massive MIMO makes the user's signal-to-interference-and-noise ratio (SINR) depend not on small-scale fading, but on large-scale fading only \cite{Marzetta:2010}. Thus the user peak rates can be predicted {\it a priori} and simple near-optimal scheduling policies can be designed that have lower overheads than their conventional MIMO counterparts \cite{Bethanabhotla_Bursalioglu2014}.

The mmWave band is expected to play a key role in 5G. Indeed, the mmWave band can offer orders of magnitude additional available bandwidth with respect to existing cellular networks \cite{Rappaport2011}. Similar to conventional MIMO in LTE, one of the most critical challenges in networks operating on mmWave bands is the overheads for CSI acquisition at the BS. The traditional CSI-acquisition  approaches employed in FDD-based LTE rely on reference signaling in the downlink and subsequent CSI feedback through uplink. The inherent use of massive MIMO on mmWave bands makes the FDD-based CSI learning and feedback overwhelming. One exception is  Joint Spatial Division and Multiplexing (JSDM)  \cite{JSDM_mmWave}.

In contrast, the TDD operation allows learning the downlink CSI ``fast" via uplink (UL) training  and by exploiting  UL/DL channel reciprocity. Since the channel coherence time is inversely proportional to the carrier frequency \cite{Rappaport:2002}, such TDD-based operation is even more attractive in mmWave bands. First, CSI on mmWave bands decorrelates much faster than in the bands used by LTE-base cellular systems. In addition, high levels of shadowing caused by the appearance of obstacles also leads to more dramatic swings in the path losses.

Increasing the carrier frequency from the LTE bands to mmWave bands results in a severe increase in path-loss \cite{Rappaport:2002}. The ability to pack a large number of antennas into a small footprint at mmWave enables pathloss compensation via large antenna array gains. Although massive-MIMO type beamforming gains can increase the mmWave cell coverage area, mmWave cells are inherently expected to be deployed as small cells. The combination of harsh and rapidly changing channel characteristics, however, imply that mmWave small cells will not be able to provide adequate coverage. Indeed, a significant fraction of user terminals is expected to be in outage in mmWave bands and thus must be supported over e.g., the LTE fabric. This renders the need for 5G heterogeneous networks that comprise of multi-tier networks operating over a broad range of frequency bands. As a result, it is reasonable to optimize the mmWave band network so as to maximize the throughput per unit area it can provide to the users that it can serve, without requiring the mmWave band network to provide adequate coverage on its own. At the same time, the preceding argument implies that the connectivity in mmWave small cells is highly intermittent and communication needs to be rapidly adaptable. Thus, an important problem addressed by our work is the following: from the network perspective, how do the BSs detect users {\em fast} in a TDD-based network?

\begin{figure}
\centering
\includegraphics[width=0.35 \textwidth]{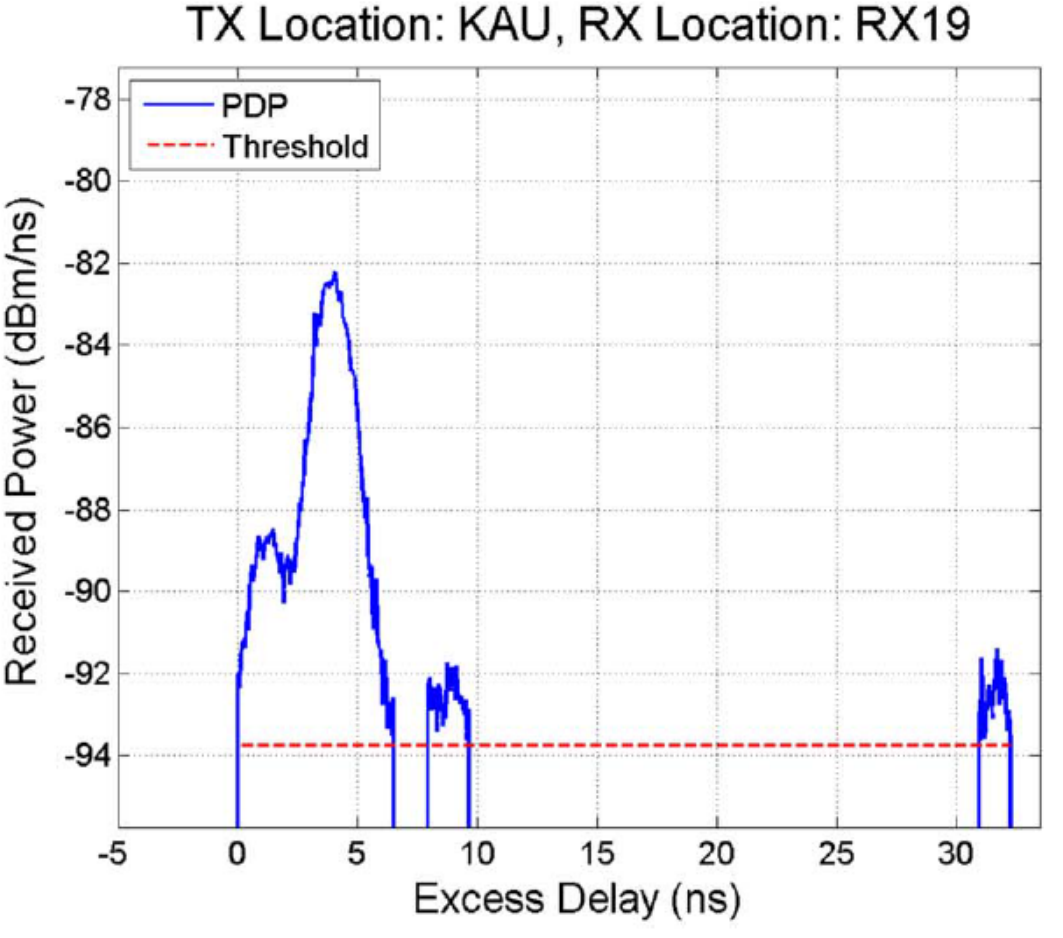}\vspace{-0.1in}
\caption{The channel is sparse in the time domain \cite{Rangan:2014}}
\label{multipath}
\end{figure}

At the same time, it is important that CSI acquisition is as efficient as possible. As in mmWave bands, the channels are sparse, i.e., they have very few dominant multipath components. As shown in \cite{Bajwa:2010, Berger:2010}, compressed sensing can exploit the channel sparsity (see Fig. \ref{multipath} where there are only 4 paths in the range of 35ns) and can harvest large gains in the pilots dimensions per user needed for channel estimation. With compressed sensing, the required number of pilot dimensions for training a user scales linearly (up to a log-factor) with the number of multi-paths $S$, contrary to the traditional approach which relies on the product of the bandwidth $W$ and the delay spread $\tau_{max}$ \cite{Falconer2002}. Clearly, when the channel is sparse, i.e., $S\ll W\tau_{max}$, pilot dimensions  per user can be saved in the training phase,  allowing more resources for data transmission and for training additional user channels. In addition, when mmWave massive MIMO offers the ability to resolve multiple Angles of Arrival (AoAs) and Angles of Departure (AoDs), which indicates that the channels become even {\em  sparser} in the AOA/AOD/delay domains \cite{Rappaport_3D, Love_mmWave, Bajwa:2010}. These properties motivate the application of compressed sensing in channel estimation in mmWave networks such as \cite{Love_mmWave}.

In this paper, we consider a small cell network where massive MIMO is operated on mmWave bands. In particular, we leverage a rudimentary binary pilot-code design combined with compressed sensing-based channel estimation and aggressive pilot reuse. As a result, a novel compressed sensing-based pilot assignment and reuse for mobile users in mmWave cellular systems is advocated, which allows a dramatic increase in the number of users simultaneously supported by the entire network. Subsequently, the system multiplexing gains and multiplexing gains per unit area can be significantly boosted compared to  traditional approaches that do not exploit such aggressive pilot reuse.

\section{System model} \label{sec_sys}

\subsection{Channel Model}

Consider a wireless cellular network where there are $N$ BSs, each equipped with $M_{BS}$ antenna elements, and $K$ UEs, each equipped with $M_{UE}$ antenna elements. We assume that the network is operated in the TDD mode. Due to the channel reciprocity in the TDD system, CSI is learned at the BS in the uplink training phase. For a uniform linear antenna array (ULA) with critically spaced antenna elements, the BS is able to resolve $M_{BS}$ AoAs and the UE is able to resolve $M_{UE}$ AoDs via linear transformation. Given a pair of a specific AoA and a specific AoD, the channel on the mmWave band is usually very {\em sparse}, i.e., the channel consists of very few significant propagation paths. To see this, let us assume that the signaling bandwidth is $W$, i.e., the time resolution is $\frac{1}{W}$. The sampled channel between a pair of AoA and AoD consists of $S$ significant channel taps, each with the gain $\beta_s$ and the delay $\tau_s,~s = 1,\dots, S$. Suppose that the delay spread of the channel is $\tau_{\max}$ and thus the channel has a total of $W\tau_{\max}$ taps (without loss of generality, we assume that $W\tau_{\max}$ is an integer). The channel sparsity implies that the number of significant channel taps $S \ll W \tau_{\max}$. Beyond the traditional frequency domain equalization channel estimation method, the sparse property motivates the investigation on compressed sensing-based channel estimation methods. Therefore, the channel estimation problem boils down to estimating the number of significant taps, their corresponding gains and delays.

\subsection{OFDM Signaling}

Let us consider a mmWave cellular network with OFDM based signaling. Suppose that the OFDM subcarrier symbol duration is $T$ (or the subcarrier spacing is $\frac{1}{T}$), where $T$ is usually much larger than the channel delay spread for combating intersymbol interference (ISI), say $T = 10 \tau_{\max}$. Since the signaling bandwidth is $W$, the total number of subcarriers is $WT$ and the set of subcarrier waveforms is given by $\left\{\exp \left(j 2 \pi \frac{i}{T}t \right), i = 0,1,\dots,WT-1\right\}$, where $j=\sqrt{-1}$. With OFDM signaling, the traditional frequency domain equalization channel estimation can be accomplished by first pilot-based training and then frequency domain equalization. Given the coherent bandwidth $\frac{1}{\tau_{\max}}$, the total number of pilots we need is thus $W \tau_{\max}$ when assigning a pilot for each coherent bandwidth. However, by  exploiting the channel sparsity via compressed sensing, we can reduce the required number of pilots for channel estimation.

\section{Energy based Binary Detection}

In this section, we present a simple method for detecting if a UE is in the vicinity of a BS, based on the observation of the signal received at that BS. This method builds a part of the foundation of this work.

Consider the uplink transmission from one UE to one BS. By assuming the BS is equipped with $M_{BS}$ antennas and the UE is equipped with one antenna, the received $M_{BS} \times 1$ signal vector ${\bf y}$ at the BS is given by
\begin{eqnarray}
{\bf y}=\sqrt{g}{\bf h} x +{\bf z}=\sqrt{gP}{\bf h}s +{\bf z}
\end{eqnarray}
where $g$ represents the large-scale path-loss from the UE to the BS, ${\bf h}$ denotes the small scale fading and each entry follows from $\mathcal{CN}(0,1)$ (may not be independent), $x=\sqrt{P}s$ denotes the signal from the UE and satisfies the power constraint $\mathbb{E}[\|x\|^2]\leq P$, $s$ is the transmitted symbol normalized by power $P$, and ${\bf z}\sim \mathcal{CN}({\bf 0},{\bf I})$ is the AWGN. Then the energy of the received signal, normalized by $M_{BS}$, can be written as
\begin{eqnarray}
E^y &\!\!\!\!= \!\!\!\!& \frac{1}{M_{BS}}{\bf y}^H{\bf y}=\frac{1}{M_{BS}}(\sqrt{gP}{\bf h}s +{\bf z})^H(\sqrt{gP}{\bf h}s +{\bf z})\notag\\
&\!\!\!\!=\!\!\!\!& \frac{1}{M_{BS}}\left(gP\|{\bf h}\|^2|s|^2+2\textrm{Re}({\bf z}^H\sqrt{gP}{\bf h}s)+\|{\bf z}\|^2\right).
\end{eqnarray}
We choose the transmitted symbol from the binary set $\{0,1\}$, i.e., $s\in \{0,1\}$, so that the energy metric $E^y$ under the two options $s=0$ and $s=1$ is given by
\begin{small}
\begin{eqnarray}
E_0^y&\!\!\!\!\!\!=\!\!\!\!\!\!&E^y(s=0)=\frac{1}{M_{BS}}\|{\bf z}\|^2,\\
E_1^y&\!\!\!\!\!\!=\!\!\!\!\!\!&E^y(s=1)\!\!=\!\!\frac{1}{M_{BS}}\left(gP\|{\bf h}\|^2+2\sqrt{gP}\textrm{Re}({\bf z}^H{\bf h})+\|{\bf z}\|^2\right).\ \ \
\end{eqnarray}
\end{small}
Clearly, when $M_{BS}$ is large, $E_0^y$ representing the noise energy only will converge to its mean value $\mathbb{E}[E_0^y]=1$, and $E_1^y$ will converge to $\mathbb{E}[E_1^y]=1+gP$. Intuitively, we can estimate $s$ by calculating $E^y$ and comparing if it is significantly larger than the noise level $\mathbb{E}[E_0^y]=1$. In practice, we use the preset value $\eta \in (1,1+gP)$ as the threshold to detect
\begin{eqnarray}
\hat{s}=\left\{\begin{array}{lll}0&& E^y\leq \eta,\\1&& E^y> \eta,\end{array}\right.
\end{eqnarray}
and the error probability is given by
\begin{eqnarray}
P_e=\int_0^{\eta}p_{E_1^y}(x)dx+\int_{\eta}^{+\infty}p_{E_0^y}(x)dx.
\end{eqnarray}
Note that the random variable $E_0^y$ follows the $\mathcal{X}^2$ distribution and its the p.d.f. is given by
\begin{eqnarray}
p_{E_0^y}(x)=\frac{(1/2)^{M_{BS}}}{\Gamma(M_{BS})}(\sqrt{2M_{BS}}x)^{M-1}e^{-\sqrt{\frac{M_{BS}}{2}}x}
\end{eqnarray}
where $\Gamma()$ is the Gamma function. Also, when $gP\gg 0$, $E_1^y$ also approximately follows the $\mathcal{X}^2$ distribution and its the p.d.f. can be written as
\begin{eqnarray}
p_{E_1^y}(x)\approx \frac{(1/2)^{M_{BS}}}{\Gamma(M)}\left(\sqrt{\frac{2M_{BS}}{1+gP}}x\right)^{M-1}e^{-\sqrt{\frac{M_{BS}}{2(1+gP)}}x}.
\end{eqnarray}
The detection error probability $P_e$ can be minimized by setting the threshold $\eta$ as the solution of $p_{E_0^y}(x)=p_{E_1^y}(x)$.
In fact, it can be easily verified that $P_e\rightarrow 0$ when $M_{BS}\rightarrow +\infty$. In a network where the users have different path-losses, we can compute offline every $\eta_k$ regarding each $g_k$ where $k=1,2,\cdots,K$, and choose the minimum of the qualified $\eta_k$ values to be the targeted $\eta$, or directly use $gP$ that guarantees the minimum received SNR at the BS to calculate $\eta$ in the network wide.

In the reminder of this paper, we use $\hat{s}=1$ and $\hat{s}=0$ to denote the received signal is detected to have a high energy level and a low energy level, respectively. In addition, for simplicity, we assume $P_e=0$, which can be nearly guaranteed when massive MIMO ($M_{BS}\gg 0$) is deployed at the BS.


\section{Compressed sensing-based pilot assignment}\label{sec_cs}

\subsection{The Single User Scenario}

In compressed sensing, we randomly select $M$ pilot tones out of the $WT$ subcarriers. The number of pilot tones that we need for channel estimation depends on the channel sparsity parameter $S$. For example, to have a comparable channel estimation performance as the least square methods in frequency domain equalization, we set $M = 5S$ in this paper.

Let $\mathcal{M}$ be the set of $M$ (pseudo-) randomly selected pilot tones, where $|\mathcal{M}|=M$ and each element of $\mathcal{M}$ is selected from the set $\{1,2,\cdots,WT\}$. We denote the training signal as $x(t)$, which can be written as
\begin{equation}
x(t) = \sqrt{E} \sum_{n \in \mathcal{M}} g(t) \exp\left(j 2\pi \frac{n}{T} t\right),~0 \leq t \leq T,
\end{equation}
where $E$ is the symbol energy and $g(t)$ is the pulse shape. At the BS, the received signal (at the particular AoA) is matched-filtered with the OFDM basis waveforms $\left\{g(t) \exp\left(j 2\pi \frac{n}{T} t\right)\right\}_{n \in \mathcal{M}}$. Then, the matched-filtered outputs are collected into an $M \times 1$ vector $\mathbf{y}$, which is given by \cite{Goldsmith:2005}
\begin{equation}
\mathbf{y} = \sqrt{E} \mathbf{X} \mathbf{h}_\eta + \mathbf{z},
\end{equation}
where $\mathbf{X}$ is an $M \times W\tau_{\max}$ matrix (the so-called sensing matrix) with its rows
\begin{align}
&\left\{ \left[1, \exp\left(-j \frac{2\pi}{WT}n\right), \exp\left(-j \frac{2\pi}{WT}n2\right), \cdots \right. \right. \nonumber\\
&~~~~~~~~~~ \left. \left. \dots, \exp\left(-j \frac{2\pi}{WT}n(W\tau_{\max}-1)\right)\right] \right\}_{n \in \mathcal{M}}.
\end{align}
the $W\tau_{\max} \times 1$ vector $\mathbf{h}_\eta$ is the sampled channel with $S$ non-zero elements (corresponding to the $S$ significant channel taps), and $\mathbf{z}\sim \mathcal{CN}({\bf 0}, {\bf I})$ is an $M \times 1$ AWGN vector.

We use the Dantzig selector \cite{Candes:2007,Bajwa:2010}, which is one of the sparse signal recovery techniques that has asymptotic performance guarantee, to estimate the sampled channel $\mathbf{h}_\eta$. Thus we have the following linear programming problem:
\begin{align}
\mini_{\mathbf{h}}&~~ \|\mathbf{h}\|_1 \nonumber\\
\textrm{subject to}&~~ \|\mathbf{X}^H (\mathbf{y} - \sqrt{E}\mathbf{X}\mathbf{h})\|_{\infty} \leq \epsilon,
\end{align}
where $\|\cdot\|_1$ represents the 1-norm, $\|\cdot\|_{\infty}$ represents the infinity norm, and $\epsilon$ is a system parameter that can be chosen to control the error.

Finally, we emphasize that the gain of the compressed-sensing based channel estimation scheme compared to traditional frequency domain equalization is the reduction of the number of pilot tones for uplink training, from $W \tau_{\max}$ to $M$.

\subsection{The Multiple User Scenario}

In this subsection, we extend the scheme introduced above to support concurrent uplink training for multiple UEs. Suppose there are $K$ UEs in the system. If $K M \leq W T$, we can pre-allocate $K$ orthogonal pilot sequences, each consisting of $M$ non-overlapping pilot tones/subcarriers, for the $K$ UEs. The pre-allocation proceeds as follows: We first pseudo-randomly choose a pilot sequence (denoted as $\mathcal{M}_1$) for UE 1. Then, we pseudo-randomly choose another pilot sequence (denoted as $\mathcal{M}_2$) from the remaining pilot tones $\{1,2,\dots, WT\}\setminus \mathcal{M}_1$ for UE 2. Repeat this procedure by choosing tones for a new UE from the remaining tones excluding those that have been chosen, until we finally obtain all $K$ pilot sequences  $\mathcal{M}_i,~i = 1,\dots,K$, where $\mathcal{M}_i \subset \{1,2,\dots, WT\}, |\mathcal{M}_i| = M, \mathcal{M}_i \cap \mathcal{M}_j = \emptyset, \forall i \neq j$.
The $K$ pilot sequences are pre-stored at the BS and pre-distributed to the UEs.

Now, all UEs simultaneously transmit training signals at their dedicated pilot tones. The BS can estimate the sparse channel for UE $i$ by matched-filtering the received signal with UE $i$'s corresponding set of subcarrier waveforms $\left\{g(t) \exp \left(j 2\pi \frac{n}{T} t \right)\right\}_{n \in \mathcal{M}_i},~i=1,\dots,K$.

We denote by $\rho$ the number of UEs that can be simultaneously supported by the system for uplink training, which is the performance metric quantifying the performance of different pilot assignment schemes. Thus, for the traditional frequency domain equalization scheme $\rho_{FQ} =\frac{WT}{W\tau_{\max}}$, and for the compressed sensing-based scheme, we have $\rho_{CS} = \frac{WT}{M}$.

\section{Aggressive pilot reuse} \label{sec_apl}

In a cellular network, to reduce the amount of resources allocated to uplink training, we can reuse the same pilot tones for uplink training when the UEs are located far away from each other (to combat with pilot contamination \cite{Jose:2011}). Due to the UE mobility, the reuse distance for mobile UEs, denoted as $R_{\textrm{mobile}}$, can be much larger than that for static UEs, denoted as $R_{\textrm{static}}$, and thus it would reduce the system efficiency.

Here we propose an aggressive pilot reuse scheme for mobile UEs, in which we reuse the pilot tones for mobile UEs at a shorter distance, say at $R_{\textrm{static}}$, the same as static UEs. On the one hand, such aggressive reuse increases the efficiency; on the other hand, it may cause collisions or pilot contamination due to the mobility of the UEs. Collisions occur when two UEs that are close to each other use the same pilot tones. For example, suppose that UE A and UE B use the same pilot tones and they are separated by a distance of $R > R_{\textrm{static}}$. However, at the next time slot, if they move towards each other such that $R < R_{\textrm{static}}$, a collision occurs.

Thus, to enable aggressive pilot reuse, and due to the properties of mmWave channels that we introduced in the introduction, we must design proper ``pilot sequences'' to {\em very fast} detect collisions and identify UEs if there is no collision.

\subsection{Pilot Sequence Design}

Let us first introduce the concept of a UE group. A UE group is defined as a set of UEs that are using the same pilot dimensions. Note that here we use the phrase ``pilot dimensions" rather than ``pilot tones" since a UE may not be active at all pilot dimensions. Then we assign all the UEs in the same group with \textit{overlapping} pilot sequences chosen from the same pilot dimensions.
For the pilot sequences to be assigned to UEs in a UE group, we design them to satisfy the following four criteria: given the received signal at the BS,
\begin{enumerate}
\item if there is one and only one UE nearby, the BS can detect and identify which UE it is;
\item if two or more UEs are nearby, the BS can detect the collision;
\item if there is no UE nearby, the BS can claim there is no UE to serve;
\item if there is no collision, we should have comparable channel estimation performance as the case without aggressive pilot reuse.
\end{enumerate}
The rationale behind the criteria 1) and 3) is that each BS only needs to serve a small subset of UEs due to the cell densification, and for simplicity here we assume that at each time every BS {\em simultaneously} serves up to only one UE.

We consider the following design of pilot sequences. Let us first fix a UE group with $K_G$ UEs. Suppose that the UE group is assigned with $L$ pilot dimensions (subcarriers) where $L = W\tau_{\max} + 1$ when using frequency domain equalization recovery and $L = M + 1$ when using compressed sensing-based recovery. Without loss of generality, we assume that $L \geq K_G$. We consider the following $L \times L$ codebook $\mathcal{C}$ for the pilot sequences design (Note that the code design we show here is only an example to present the insight of detecting and identifying UEs in mmWave networks):
\begin{align}\label{eqn:code1}
 \mathcal{C} \triangleq \left[ \begin{array}{ccccc}
 1 & 1 & \cdots & 1 & 0 \\
 1 & 1 & \cdots & 0 & 1 \\
 \vdots & \vdots & \cdots & \vdots & \vdots \\
 1 & 0 & \cdots & 1 & 1 \\
 0 & 1 & \cdots & 1 & 1
 \end{array} \right].
\end{align}
In particular, each column of $\mathcal{C}$ corresponds to the pilot sequence for UE $i,~i = 1,\cdots,K_G$. For each UE $i$ in the UE group, UE $i$ transmits pilots in the pilot dimensions indicated by a ``1'' (high energy level) and does not transmit in the pilot dimensions indicated by a ``0'' (low energy level).

We claim that the pilot sequences design $\mathcal{C}$ satisfies the four criteria above and thus can support up to $L$ UEs. To see this, let us verify each criteria sequentially:

(1) When there is one and only one UE nearby, the BS will detect the signal with low energy level at one of the $L$ pilot dimensions after matched-filtering/energy-detecting the received signal. Since each pilot sequence has only one ``0'' in the $L$ dimensions, the BS can identify the corresponding UE by reading the position of the low energy level.

(2) When there is a collision, the BS will see the sum of signals carrying the pilot sequences from at least two UEs in the UE group. After matched-filtering/energy-detecting the received signal, the BS can see all the pilot dimensions have high energy levels, indicating that a collision occurs (because the vector with all ``1''s does not belong to any column of $\mathcal{C}$);

(3) When no UE is nearby, the BS will see that all the $L$ dimensions have low energy level, i.e., the noise level.

(4) When there is no collision, the total number of pilot tones that are used by a UE (the number of ``1''s in the pilot sequence) is $L -1 = W \tau_{\max}$ or $M$. As a result, we have a comparable performance as the case without aggressive pilot reuse for the frequency domain equalization or the compressed sensing-based recovery.

Extensions of the code design in (\ref{eqn:code1}) can be developed so that given the total number of pilot dimensions $L'$ per user for estimating its channel, the system is able to support $K>L$ users by adding more that one $0'$s in each column of (\ref{eqn:code1}). One such family of codes that includes the code in (\ref{eqn:code1}) is parameterized by a pair of positive integers $L'$ and $l$ where $L=L'+l$, $L'$ and $l$ represent the number of $1'$s and $0'$s in the code sequence design, respectively. Similar to the procedure above, it can be easily verified that the four criteria of the code design are also satisfied, by replacing with ``only $l$ 0's" in the $L$ dimensions" in criteria (1), ``less than $l$ 0's" in criteria (2) and ``$L-l=M$" in criteria (4). Thus, given an integer number $L$, a total number of $K\leq K^l_{\textrm max}=\binom{L'+l}{l}$ users can be supported, at the cost of code efficiency reducing from $\frac{L'}{L'+1}$ to $\eta(l)=\frac{L'}{L'+l}$. In fact, given the value of $L'$, there is a tradeoff between $K^l_{\textrm max}$ and $\eta(l)$ since they cannot increase simultaneously. In our study, to achieve the best tradeoff, given the total number of users $K$ and after fixing the value of $L'$, we choose the value of $l$ so that $K^{l-1}_{\textrm max}<K\leq K^l_{\textrm max}$, and the corresponding code spectral efficiency is given by $\eta(l)=\frac{L'}{L'+l}$.

We define the collision probability, denoted by $p$, as the ratio of the average number of UEs in collisions and the total number of UEs in the network. The value of $p$ depends on the network topology and the UE mobility pattern.

\subsection{A Case Study} \label{sec_case}

\begin{figure}
\centering
\includegraphics[width=0.4\textwidth]{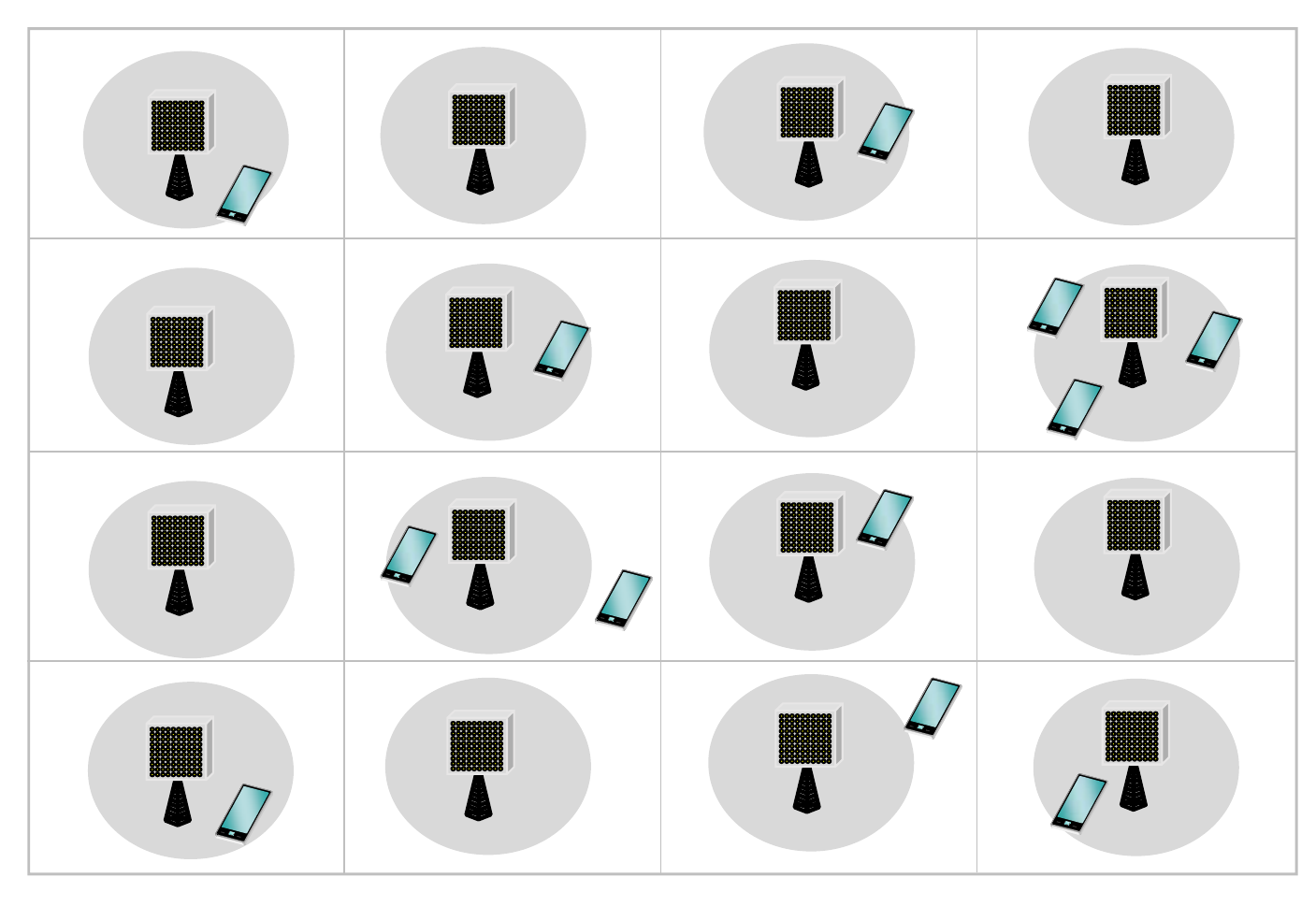}\vspace{-0.1in}
\caption{Network topology, $N = 16$ cells, $K_G = 12$ UEs.}
\label{fig2}
\end{figure}

To see how the new code design works, let us consider a simple case. We consider a specific network topology and the mobility model to compute the collision probability under the aggressive pilot reuse scheme above. As shown in Fig.~\ref{fig2}, we assume that there are $N$ cells represented by the grey colored circles, each with one BS. Note that due to the large path-losses and high level shadowing of mmWave channels, the grey colored cellular regimes may not fully cover the entire network, which means that the UEs falling outside the grey colored regime are in outage. In this case, the UEs in outage can still be served in other bands by the HetNet, because mmWave is viewed only an option to increase the spectral efficiency, and we do not need the BSs to serve {\em all} UEs on mmWave bands. In Fig.~\ref{fig2}, the outage probability denoted by $p_{out}$ can be calculated as the ratio of area of a rectangle excluding its cellular regime and the area of that rectangle. Also, although there is only one UE group with $K_G$ UEs in Fig.~\ref{fig2}, it is straightforward to generalize to the case with multiple UE groups on different bands, as what we do in Section~\ref{sec_num}.). To model the mobility of the UEs, we assume that the UEs are distributed uniformly at random among the entire network, i.e., the $N$ rectangle regimes. Thus, the probability of a UE {\em not} in outage is given by $\alpha=1-p_{out}$. Furthermore, we consider a noise-limited mmWave system \cite{Rangan:2014}, in which the received SNR within a cell is high enough for communication and there inter-cell interference is under the noise level, i.e., inter-cell interference is neglected. As a result, when two or more UEs are located in the same cell, there is a collision (since they belong to the same UE group reusing the same pilot dimensions). For example, in Fig.~\ref{fig2} there are 3 UEs in collision in one of the 16 cells, and 2 UEs in outage.

Next, we proceed to compute the expected number of cells that have exactly one UE. We define the indicator random variable $X_i$, where $X_i = 1$ if the $i$-th BS cell (grey colored circle) contains exactly one UE and $X_i = 0$ otherwise. As a result, the total number of cells that have exactly one UE is given by $X = \sum_{i=1}^N X_i$. Then we have
\begin{align}
\mathbb{E}[X] &= \sum_{i=1}^N \mathbb{E}[X_i]=\sum_{i=1}^N \Pr(\textrm{i-th cell contains exactly 1 UE}) \nonumber\\
&=N \binom{K_G}{1} \left(\frac{\alpha}{N} \right) \left(1-\frac{\alpha}{N} \right)^{K_G-1} \nonumber\\
& =\alpha K_G\left(1-\frac{\alpha}{N} \right)^{K_G-1}.
\end{align}
Since $\mathbb{E}[X]$ is the average number of UEs that do not collide with any other, the collision probability is thus given by
\begin{eqnarray}
p = \frac{K_G-\mathbb{E}[X]}{K_G} =1 - \alpha\left(1-\frac{\alpha}{N} \right)^{K_G-1}.
\end{eqnarray}
Based on the expressions for $\mathbb{E}[X]$ and $p$, we have a number of observations. First, if $N$ is large but $K_G$ is fixed, then we have $\mathbb{E}[X]\approx \alpha K_G$ and $p\approx 1-\alpha=p_{out}$, which implies that cell densification increases the degrees of freedom of the network. Second, if $K_G$ is large but $N$ is fixed, then $\mathbb{E}[X]\approx 0$ and $p\approx 1$, which implies that increasing the number of UEs will cause more collisions of UEs. Finally, when $N$ is large, $\mathbb{E}[X]$ is maximized when $K_G^* \approx N/\alpha$. To see this, we take the derivative of $\mathbb{E}[X]$ with respect to $K_G$ and let
\begin{equation}
\frac{d\left(\alpha K_G\left(1-\frac{\alpha}{N} \right)^{K_G-1}\right)}{dK_G}=0\Longrightarrow K_G^* \approx \frac{N}{\alpha}.\label{eqn:optKG}
\end{equation}
Under this case, the collision probability (ratio) is $p \approx 1-\alpha/e$. Furthermore, we denote by $\frac{WT}{L} K_G (1-p)$ the number of UEs that can be simultaneously supported by the system with aggressive pilot reuse for uplink training, the product of the number of UE groups, the number of UEs in a UE group, and the non-collision probability $1-p$. For frequency domain equalization and compressed sensing-based recovery, we have
\begin{small}
\begin{eqnarray}
\rho_{AG-FQ}&\!\!\!\!=\!\!\!\!& \frac{WTK_G (1-p)}{W\tau_{\max}+1} =\frac{WTK_G\alpha}{W\tau_{\max}+1}\left(1-\frac{\alpha}{N} \right)^{K_G-1}.\label{eqn:ag_fq}\\
\rho_{AG-CS}&\!\!\!\!=\!\!\!\!&\frac{WTK_G (1-p)}{M+1} =\frac{WTK_G\alpha}{M+1}\left(1-\frac{\alpha}{N} \right)^{K_G-1}.\label{eqn:ag_cs}
\end{eqnarray}
\end{small}
In addition, we rewrite $\rho_{FQ}$ and $\rho_{CS}$ introduced in Section \ref{sec_cs}.B by incorporating the outage probability. That is,
\begin{eqnarray}
\rho_{FQ} &\!\!\!\!=\!\!\!\!& WT\alpha/(W\tau_{\max}),\label{eqn:fq}\\
\rho_{CS} &\!\!\!\!=\!\!\!\!& WT\alpha/M.\label{eqn:cs}
\end{eqnarray}
Finally, we note that the complexity of the compressed sensing based channel recovery depends on how the corresponding linear program is solved \cite{Berger:2010}.

\section{Numerical Results}\label{sec_num}

In this section, we evaluate the performance of the proposed  aggressive pilot reuse mechanism for uplink training and user detection via numerical simulations. We assume that the total number of subcarriers is $WT = 1000$. The number of channel taps is $W \tau_{\max} =100$, and there are $S = 4$ significant taps, as suggested by the measurement results in the 28 GHz and 38 GHz mm-Wave bands~\cite{Sulyman:2014}. The compressed sensing ratio is set to be 5 \cite{Bajwa:2010}, so we have $M = 5S = 20$. We consider the system topology as described in Section~\ref{sec_case} with $N = 16$, and we assume that the outage probability $p_{out}=0$ and $p_{out}=30\%$. The number of UEs in a UE group $K_G$ varies. We study the performance metric $\rho$, i.e., the number of UEs that can be simultaneously supported by the system for uplink training, under the following four schemes: frequency domain equalization $\rho_{FQ}$ in (\ref{eqn:fq}), frequency domain equalization with aggressive pilot reuse $\rho_{AG-FQ}$ in (\ref{eqn:ag_fq}), compressed sensing based recovery $\rho_{CS}$ in (\ref{eqn:cs}), and compressed sensing based recovery with aggressive pilot reuse $\rho_{AG-CS}$ in (\ref{eqn:ag_cs}).

\begin{figure}[t!]
\centering
\includegraphics[width=0.5\textwidth]{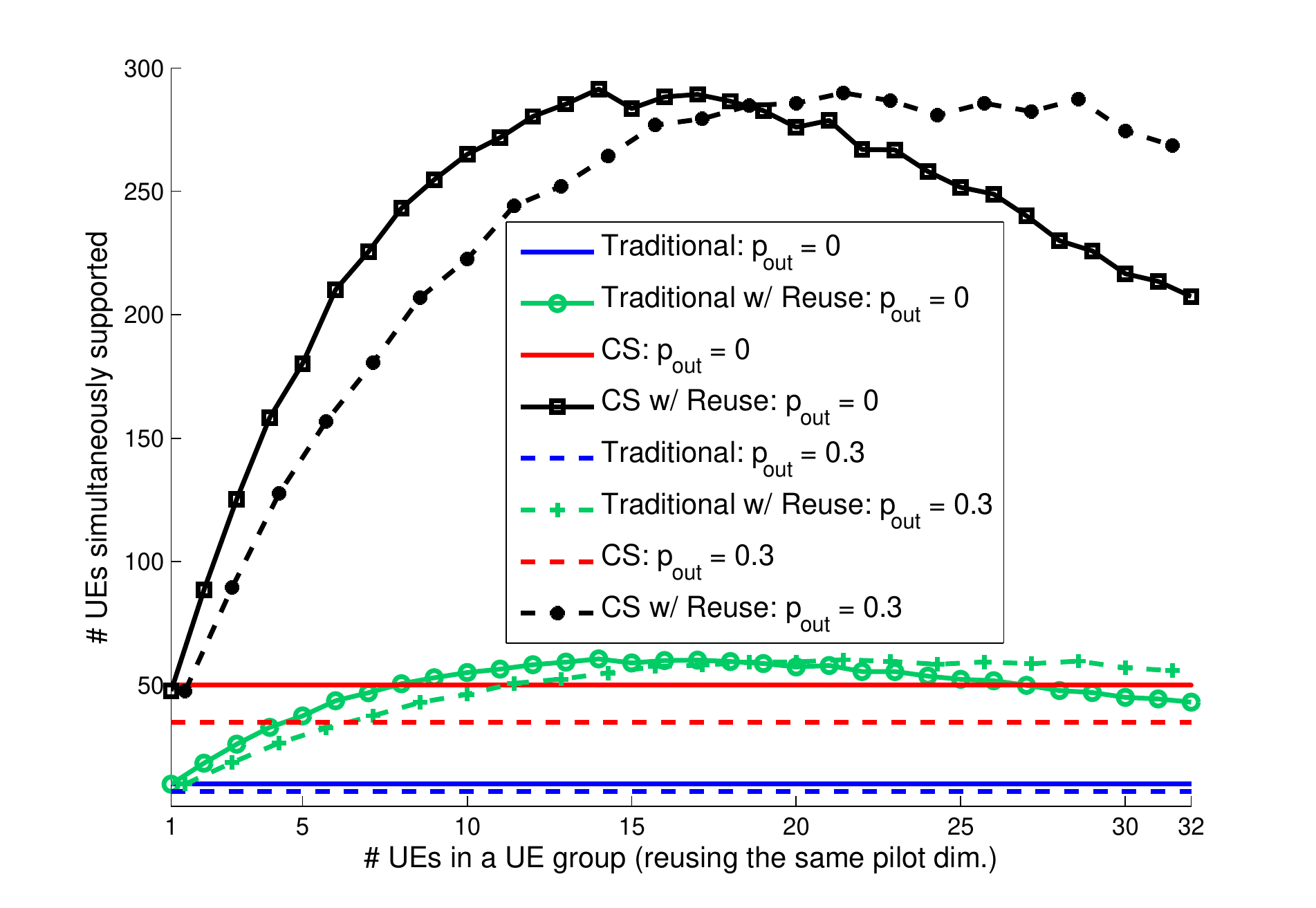}
\caption{Performance comparisons of the four schemes}
\label{fig1}
\end{figure}

The numerical results are shown in Fig.~\ref{fig1}, where the blue, green, red and black colored curves represent the performance metric $\rho_{FQ}$, $\rho_{AG-FQ}$, $\rho_{CS}$ and $\rho_{AG-CS}$, respectively, and for each color, the solid and dashed curves represent the setting $p_{out}=0$ and $p_{out}=0.3$, respectively. From Fig.~\ref{fig1}, we use the $p_{out}=0$ setting (the solid curves) to present our following three interesting observations.

First, comparison of the red and blue colored curves (independent of the number of UEs in each group) reveals the performance gain harvested by the compressed sensing scheme, which exploits the wireless channel sparsity.

Second, comparison of green and blue curves reveals the performance gains due to aggressive pilot reuse. The benefits of aggressive pilot reuse can also be observed for CS based channel estimation (from the red to the black colored curves).

Next, it can be seen that the proposed ``CS w/Reuse" scheme $\rho_{AG-CS}$ (black colored curve), which exploits both channel sparsity and aggressive pilot reuse, outperforms all the other schemes. Also, when the number of UEs in a UE group, i.e., the parameter $K_G$, keeps increasing, $\rho_{AG-CS}$ first increases and then decreases, and the optimal performance is achieved at $K_G^* \approx N/\alpha$ for $\alpha=1$ when $p_{out}=0$, which is consistent with our analytical result shown in (\ref{eqn:optKG}).

Finally, from $p_{out}=0$ to the $p_{out}=0.3$ setting, it can be seen that the simulation results are similar except that the outage kills a fraction of UEs. However, the performance gains offered by the aggressive pilot reuse are not identical for different values of $p_{out}$. To see this, consider the performance gains offered by the aggressive pilot reuse for the CS-based scheme (from the red to the black colored curves):
\begin{eqnarray}
G(p_{out})=\frac{\rho_{AG-CS}}{\rho_{CS}}=\frac{MK_G}{M+1}\left(1-\frac{1-p_{out}}{N} \right)^{K_G-1}.
\end{eqnarray}
It can be easily verified that $G(p_{out})$ is an increasing function of $p_{out}$, i.e., more UEs in outage, the larger performance gains.

\section{Conclusion}\label{sec_con}

In this paper, we study the fast user detection and identification problem in small cell mmWave systems. In particular, we design pilot-assignment and pilot-reuse mechanisms, which rely on rudimentary user detection at the BS, compressed-sensing based channel estimation. Our  simulations reveal that when these mechanisms are combined with aggressive pilot reuse  significant multiplexing gain improvements can be harvested  compared to conventional methods relying on conventional pilot reuse.

\end{document}